# A Cognitive Ideation Support Framework using IBM Watson Services

*Completed Research paper*

**Samaa Elnagar**

Virginia Commonwealth University

elnagarsa@vcu.edu

**Kweku-Muata Osei-Bryson**

Virginia Commonwealth University

kmosei@vcu.edu

## Abstract

Ideas generation is a core activity for innovation in organizations. The creativity of the generated ideas depends not only on the knowledge retrieved from the organizations' knowledge bases, but also on the external knowledge retrieved from other resources. Unfortunately, organizations often cannot efficiently utilize the knowledge in the knowledge bases due to the limited abilities of the search and retrieval mechanisms especially when dealing with unstructured data. In this paper, we present a new cognitive support framework for ideation that uses the IBM Watson DeepQA services. IBM Watson is a *Question Answering* system which mimics human cognitive abilities to retrieve and rank information. The proposed framework is based on the *Search for Ideas in the Associative Memory* (SIAM) model to help organizations develop creative ideas through discovering new relationships between retrieved data. To evaluate the effectiveness of the proposed system, the generated ideas generated are selected and assessed using a set of established creativity criteria.

## Introduction

In the increasingly competitive marketplace, innovation is necessary for organizations not only to compete but also to survive. Innovation can be defined as "the diffusion of a new technique through the development and use of new ideas or behaviors" (Damanpour and Wischnevsky 2006). To develop creative ideas, organizations should maximize the average quality of ideas not the number of ideas (Girotra et al. 2010). Innovative ideas originate from the correct exploration and incorporation of disparate bodies of existing knowledge, including from external sources (Sun and Münster 2018). However, many organizations suffer from difficulties in utilizing the existing knowledge in their *Knowledge Management Systems (KMS)* (Damodaran and Olphert 2000) and the ambiguous representation of the captured ideas (Morente-Molinera et al. 2016).

*Question Answering (QA)* systems such as IBM Watson have emerged as intelligent systems for conveniently retrieving knowledge from the KMS. *QA* systems differ from traditional search engines which use keywords or search cues as input to the search. The input to the *QA* system is in the form of questions in a natural language and the output is candidate answers to the questions (Gupta and Gupta 2012). *IBM Watson* is a cognitive computing state-of-art *QA* system that is able to analyze huge amounts of unstructured data, and answer a wide variety of questions spanning a broad range of topics (Ferrucci 2011). *Cognitive Computing QA systems* mimic human reasoning and inference abilities in reaching the confident answers. These cognitive systems extended the machine ability of knowing to reach the level of wisdom (Craw and Aamodt 2018).

Idea Generation methodologies are classified into two primary groups: intuitive and logical. Intuitive methods use human intuition activities such as C-sketch, and fishbone diagram methods; logical methodologies use cognitive-based reasoning for the development of new ideas (Shah et al. 2003). One of these logical methodologies is the *Search for Ideas in Associative Memory (SIAM)* that depends on generating new ideas by creating new relationships out of the data retrieved from the *long term memory (LTM)* (Raaijmakers and Shiffrin 1981). However, the ideation process depends primarily on a myriad of influences such as group personalities, group size, in addition to organizational culture and environment. However, these issues are beyond the scope of our research (Dennis et al. 1999). Therefore, the basic assumption here is that the group performing the ideation process has no organizational or social obstructions to the ideation process such as evaluation apprehension (Diehl and Stroebe 1991).

QA systems could interactively help to generate innovative ideas, as the answers provided by QA systems can stimulate the individuals' memory to generate new connections to a new idea. Therefore, this paper aims to develop a cognitive ideation support framework that could help the organization generate innovative ideas by using knowledge from the organization's KMS and external sources. Our framework will mimic the *SIAM* cognitive ideation model and has two roles as in Table 1. The first is to stimulate ideas of the participants by actively answering their questions. The second is to create relationships from the knowledge retrieved and combine concepts to generate new ideas. The best ideas are selected and evaluated through predefined criteria such as novelty, usefulness, etc. (Morente-Molinera et al. 2016).

**Table 1. The cognitive ideation framework summary**

| Research questions | - How can simulating the *SIAM* model through Watson improve the ideation process?<br>- How can Watson maximize the retrieved knowledge from discrete body of knowledge and generate relations between retrieved concepts? |
|---|---|
| Aim | the framework mimics the idea generation process of human mind and maximize the use of knowledge from different sources to help organization generating creative ideas. |
| Architecture | The framework is based on the *SIAM* model for generating ideas. The Watson is used for stimulating ideas by answering participants questions and to retrieve knowledge from KMS |

## *Ideation and the ultimate benefit of knowledge*

Is the knowledge stored in an organization's KMS always sufficient for the generation of new ideas? The answer to this question is "No" for even if the KMS has been used ultimately the success of the ideation process is directly related to the knowledge gained from different types of knowledge sources (Morente-Molinera et al. 2016). Knowledge is stored in the KMS in many forms: relational database, blogs, forums, mind maps, discussion sessions, and reports (Kwan and Balasubramanian 2003; Farr et al. 2003). Therefore, one fundamental goal of a creative ideation process is to maximize the knowledge gained from the KMS and other external knowledge sources since maximizing the amount of relevant knowledge retrieved, maximizes the probability of finding the correct answers to a problem (Moschitti et al. 2017).

Ruchter (2003) represented knowledge in four structure formats: the "*known known*", "*unknown known*", "*unknown unknown*", and "*known unknown*". We build on this representation and define the "complete picture of knowledge" as shown in Figure 1. The "*known known*" is the data that the organization knows that it knows which is the data that could be retrieved from the KMS. The "*unknown knowns*" quadrant is the data that the organization does not know it knows or the data that cannot be reached in the KMS because of the limited capability of the search and retrieval system. *QA* systems provide better search methodology that should overcome the "*unknown knowns*". The third quadrant is the "*known unknowns*" or the data that the organization knows it doesn't know. Therefore, the organization must include data from external sources to extend its knowledge to the "missing" knowledge. The fourth quarter is the "*unknown unknowns*" or the data the organization doesn't know it doesn't know. This data is induced and it can be only obtained through ideation or innovation. Based on this representation we can conclude that a new idea is simply revealing the "*unknown unknowns*" or using the *knowns* to expose the *unknown*. Therefore, one of the main goals for building our ideation support framework to maximize the knowledge gained especially knowledge from the "*unknowns*", to maximize the ideation process efficiency.

## *The Cognitive Idea Generation Process*

Idea generation is an indispensable process to create novel products and services. Selecting the mental activities the techniques, the instructions, and the tools used in the ideation process are all critical factors to the quality of the generated ideas (Spohrer and Banavar 2015;). Based on the SIAM model for ideation, the ideation process is ignited by a stimulus is received through the human senses. This stimulus produces search cues in the Working Memory (WM) (Raaijmakers and Shiffrin 1981). Then, a search cue activates the knowledge from the individual's Long-Term Memory (LTM) in the form of an image. "Image" here means the knowledge structure that consists of *a central concept* and *relations* with other concepts. For example: the image of a "Car" is associated with gas, insurance, etc. Images have no definite boundaries but they overlap with other images (Nijstad and Stroebe 2006).The retrieved image is temporarily stored in the WM where the concepts are transformed into active knowledge.

Afterwards, ideas are generated through creating new associations between concepts or extending the active knowledge to the domain of interest. Therefore, retrieving more images and adding more concepts to the active knowledge, increasing the likelihood to generate more relevant ideas. Hence, the three central elements in the ideation process are the *stimulus,* the *search ques* and the *active image* retrieved.

The more the individuals are stimulated by different stimuli, the more search cues are generated, the more concepts retrieved to the WM, the more relations created between different concepts, and the more ideas incorporated. In addition, openness to external sources is highly correlated with the ideation performance because it stimulates inactive areas of brain and increases alertness. However, a compromise between internal and external knowledge is needed as the openness to external sources might have negative returns on the relevancy and applicability of generated ideas (Salter et al. 2015).

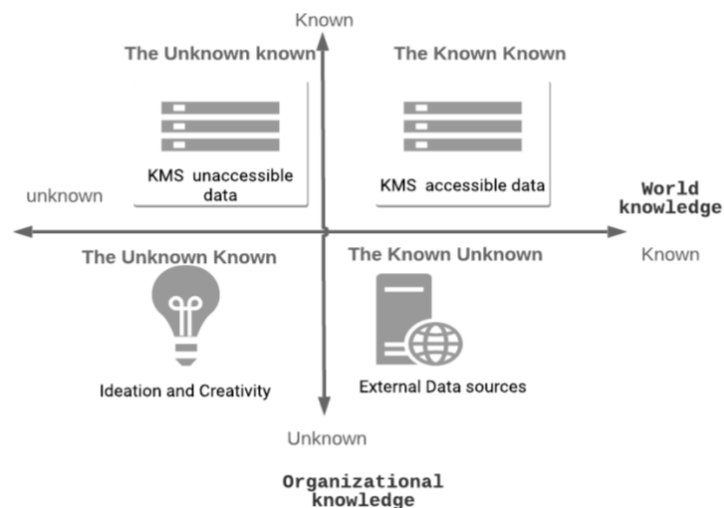
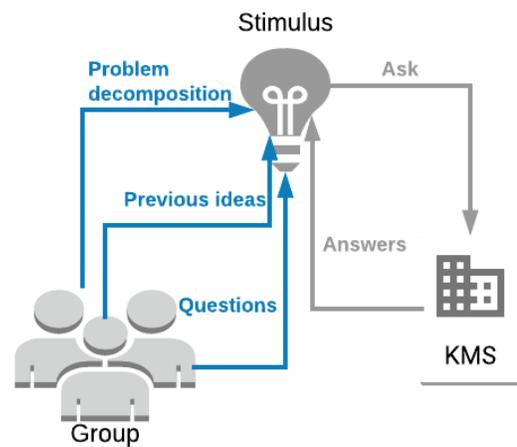

**Figure 1: The complete picture of knowledge**     **Figure 2: different ideas Stimulus**

**The limitations of individual and group ideation process**

The human brain is a rich repository of knowledge. However, the ideation process of an individual is *limited, incomplete, proportional* and *probabilistic*. The individual ideation process is *limited* because it depends on the amount of knowledge exists in individual's LTM. Even if the LTM of an individual is rich with knowledge, the image that is activated from the LTM is constrained by "*the limited attention span*" phenomena which means that only the most frequently visited and the easily accessed areas of the memory surface will be activated (Findley 2005). Therefore, the retrieved image might be *incomplete*, or doesn't not provide enough concepts to generate new ideas. the ideation process of an individual is *probabilistic* because it is hard to predict if the image retrieved will help generating new ideas (Smith 2003). In addition, the ideation process is dependent on the individual's intelligence because intelligence is related to the working memory (WM). The more information that can be held in WM at a given time, the more active knowledge can be created. That's make the ideation process proportional to the individual's WM or intelligence (Colom et al. 2008).

While the group ideation process extends the amount of active knowledge retrieved to generate creative ideas, this active knowledge can be considered less than the sum of each individual's active knowledge(Girotra et al. 2010). Most of the tacit knowledge gained by individuals in the same organization are the same, so the pool of ideas is limited. To overcome the limitations discussed earlier, the active knowledge needed to generate creative ideas must be expanded beyond the limited knowledge of the organization. Moreover, the active knowledge used in generating creative ideas must be independent of individual's WM capacity, and retrieved with degree of certainty (Smith 2003).

*Question Answering System*

The most common two types of ideas stimulus are previously generated ideas and decomposition of the problem to elements (Nijstad and Stroebe 2006). Previously generated ideas try to find specific solutions to the problem and they provide low-level tangible search cues while problem decomposition provides high-level abstract cues. However, problem decomposition is found to be significantly more beneficial than stimulus ideas in the ideation process (Luo and Toubia 2015). In addition, questions have been used as successful stimulus in ideas generation (Cooper 2008). Questions help to correct wrong beliefs and enhance the understanding of the situation. The different types of stimulus used during the ideation process is shown in Figure 2.

*QA* systems use questions in natural language format instead of keywords. Some *QA* systems can be tailored to a given knowledge domain, and they are able to query information in various formats including structured and semi-structured data (Bouziane et al. 2015). *QA* systems can be very valuable as a search tool in KMS because finding the exact information in a large amount of heterogeneous data is a complex and expensive task. *QA* as a task can be divided into three main distinct subtasks, which are: *Question Analysis, Information Retrieval*, and *Answer Extraction* (Lopez et al. 2011). The most important subtask is *Information Retrieval* as increasing the number of relevant documents increases the probability of finding the answer. In case that there is no relevant document, the *QA* system will return no answer.

Most QA systems are schema-based systems such as ontology-based QA systems which stores information based on a predefined ontological schema or knowledge graph (Chen et al. 2016). Some QA systems apply domain-independent methodologies without considering the available domain ontology such as *ontology mapping*, that transforms source ontologies to target ontologies (Roth and Schwarz 1997). However, these methodologies are still limited in dealing with unstructured data. Watson is a cognitive question answering (*QA*) that is able to analyze enormous volumes of unstructured data in different formats to answer a wide variety of questions. Nevertheless, the structured data coming from knowledge sources play a crucial role in making Watson successful. Watson use implicit knowledge found in *Off-the-shelf Knowledge Graphs* as a background knowledge to generalize to other domains. Watson has digested various structured ontology-based data sources such as Wikipedia DBpedia, Freebase and YAGO. These *Off-the-shelf* ontologies support Watson to find the correct answers and to use formal semantics and logical reasoning to find evidences in selecting answers.

## The proposed Cognitive Ideation Support Framework

This framework is designed to help groups generating new ideas. The framework mimics the SIAM model which comprises LTM and WM. We build the framework so that the LTM represents any knowledge sources that could be useful in the ideation process. The WM is responsible for temporarily storing the ideas, maintaining the repeated operations and retrieving data. The main layers of the framework are: The Long-Term Memory (LTM) layer, IBM Watson Services layer, the Support layer, the Working Memory (WM) layer, and the user interface as shown in Figure 3.

### *The Long-Term Memory (LTM) Layer*

This layer contains all possible knowledge sources used the ideation process which is similar to the LTM in the SIAM model. IBM Watson services must be trained priori with data sources in the LTM before the ideation process. The LTM consists of four components:

1. **KMS** role involves providing a repository that contains different types of data sources such as: blogs, notes, discussion session, relational databases, etc. The more operational and fiscal data added to the knowledge base, the more likely to get relevant information while generating ideas. However, integrating mind maps, Cognitive Maps, Knowledge Asset Road Maps, and Process Maps, are beyond the scope of the paper.

2. **External sources** which means any information could be obtained outside the organization KMS such as news feeds, RSS, articles, Reviews, etc. The more external knowledge regularly added to the framework, the better insights are gained from the market.

3. **The Ideas Repository**: which contains previously generated ideas from previous ideation sessions. These ideas could be used as stimulus for generating new ideas. Ideas are saved in an ontological form using the ontology generation module. The idea ontology structure is discussed later in details.

4. **Questions Repository**: this database contains all the questions asked to the *QA System* during the ideation session. The questions are saved and retrieved according to the context they were asked. These questions can be used as a stimulus for the group to generate ideas in similar context. Answers are also saved in the database and attributes such as Watson confidence level and users satisfaction with the answers are added.

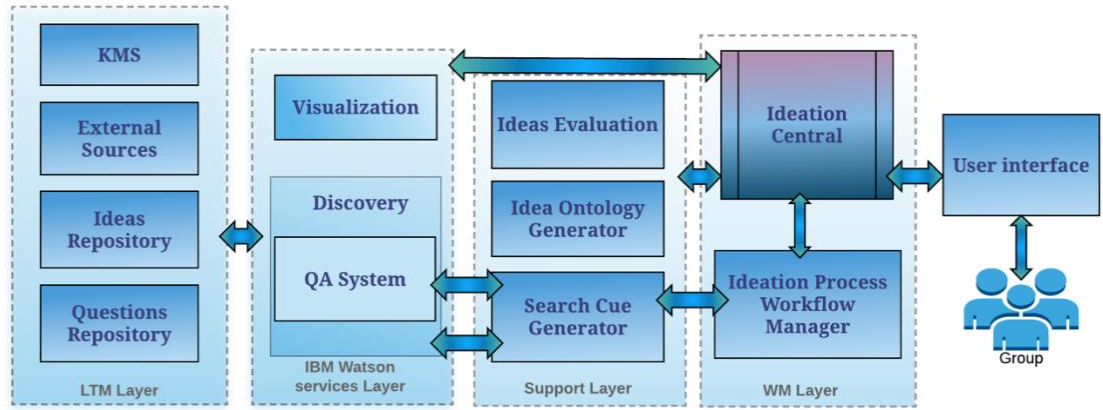

**Figure 3: Ideation Assistant Framework**

## IBM Watson Services Layer

Watson tools are available as a set of open APIs and SaaS products. Developers can build applications by combining multiple APIs(Natsev et al. 2008). In our framework, Watson services are responsible for the cognitive search, retrieval suggestion of ideas and answers. In the framework, we used four modules:

**Watson Discovery**

This service is used to find concepts, entities, sentiments and connections between different concepts in similar context. The Discovery service must process the knowledge stored in the *LTM* layer. Watson QA service is integrated within Watson Discovery to allow submitting queries using *Natural Language* question format. Answering a question involves four steps: *Question Analysis, Hypothesis Generation, Hypothesis Evidence Scoring, Final Confidence, Merging and Ranking* (Moschitti et al. 2017). *Question Analysis* is for understanding how to solve the questions. In *Hypothesis Generation, QA* produces candidate answers' snippets. *Hypothesis Evidence Scoring* considers each candidate answer a hypothesis which is given a score that represents the degree of Watson confidence in the answer. In the *Final Confidence Merging and Ranking step*, Machine-Learning techniques are used to rank the best-supported hypothesis among hundreds of candidate answers (Collinaszy et al. 2017).In our framework this service is applied to help building broader images of the problem by expanding the generated concepts and creating relationships between them as discussed Algorithm 1.

**The Visualization API**

Visualization is essential in the ideation process to enhance the understanding of the concepts generated (Tsoi et al. 2017). Watson has wide variety of runtime analytics and visualizations APIs that create holistic network maps at run time once the user makes the request. Watson visualizations extend the display of the relation between concepts, to using evidence from data to compose visual networks. Consequently , group can have a full depiction of the concepts and identify novel connections that can hardly be reached by single person (Lee et al. 2016). This step imitates the brain when it is trying to build relationships between different concepts.

## The Working Memory (WM) Layer

Extending Baddeley and Hitch (Baddeley 1997) definition of WM, the main modules of WM are:

***Central Executive***: This module is the central component in the ideation process and it has the following responsibilities:

1.It controls the interaction between the user interface and the framework.
2. It is responsible for managing activities such as creating stimulus, calling different APIs, data retrieval and organizing the idea generation activities.
3. It is used as indirect communication tool between different modules such as the communication between the idea ontology generator and ideas repository.

***Temporal Storage***: This module is responsible for temporarily storing knowledge related to the ideation process such as ideas, concepts, questions, and relations till they are saved in the framework's LTM.

---

**Algorithm 1** Stimulation Phase
**Input**: Given initial stimulus Ideas $I\{..\}$, Questions $Q\{..\}$, Concepts $C\{..\}$, Relation $R\{..\}$, Answers $A\{..\}$
**Output**: $I_d$
**Begin Round**
  *do*
    *Find context* ($q_a$)
    *Extract concepts* ($C$)
  *If Questions needed*
    *Ask Questions*($q_a$)
    *Suggest Questions* $q_s$
    *Add Questions* to $Q\{..\}$
    *Add $C_t\{..\}$ and $R_t\{..\}$*
  *else*
    *Go* to **Algorithm 2**
    *Add Concept $C_t$ to $C\{..\}$, and Relation $R_t$ to $R\{..\}$*
  *end if*
  *While* ($C\{..\}$ is not sufficient)
  *Visualize Concepts*( $C$, $R$)
  *Create Ideas* ($C$, $R$)
  *Add $I_d$ to $I\{..\}$*
**End Round**

---

**Algorithm 2** Perform QA
**Input:** asked *Questions*($q_a$)
**Output:** $C_t\{..\}$ and $R_t\{..\}$
**Begin**
  *Ask Questions $Q\{..\}$ to QA module*
  *Transform answers $A\{..\}$ to concepts $C_t\{..\}$ and relations $R_t\{..\}$*
  *Visualize Concepts ($C$, $R$)*
  *If* group approves $C_t\{..\}$ and $R_t\{..\}$
    *return $C_t\{..\}$ and $R_t\{..\}$*
  *else if group wants to continue*
    *Repeat*
  *else return*
  *end if*
**end**

---

### Ideation process Workflow Manager

Idea generation is an iterative process that has many rounds where concepts and relationships increase with each iteration. The first step in the ideation process focuses on the mission definition or problem decomposition (Blindenbach-Driessen et al. 2010; Farr et al. 2003). The process begins by stimulating the group members using different types of stimulus. Initial stimulus are the initial ideas created by the group and problem decomposition concepts. Then, the group starts to ask questions about the problem. For example, if the problem is "how to enhance an existing product to increase profit?". The initial stimulus could be "what are the drawbacks of the current product?", "How much is last year profits?". These initial questions are fed into the discovery service to get answers. The answers are retrieved in both textual and visual formats.

Extending the ideation process by Nijstad (Nijstad and Stroebe 2006) by adding QA as a stimulus, each round in the ideation has three loops: the *Stimulation phase,* the *QA phase* and *Idea creation phase*.

- Each round begins with the *Stimulation phase* where the group is trying different stimulus to produce different concepts and relations between these concepts.
- If the group needs more information, group members ask questions via the QA *System*. The *QA System* provide answers and suggests similar questions in the same context to be new stimulus.
- If the group is satisfied with the concepts and relations from the *Stimulation* phase, the *Ideation* phase starts, and the group starting coming up with new ideas. The ideas resulted from the *Ideation* phase are used as new stimulus in the next round. The ideation process workflow can be summarized in figure 4.

What is the stop criteria? the group can stop when they cannot come with more ideas or they are satisfied with the ideas they developed. Although each round has three loops, group can quit the process at any loop as long as some ideas are generated. The ideation workflow can be summarized in Algorithms 1 &2.

### *Support Layer*

This layer involves three modules that support the ideation central module in the ideation process, these modules are:

### Search Cue Generator

This module is dedicated to refine and expand insights generated during ideation session in two ways. The first is analyzing the context of each concept, idea or question asked by the group then trying to retrieve the relevant ideas and questions from the LTM as stimulus back to the group.

### Idea Evaluation Module

At the end of the ideation sessions, many ideas might be generated. However, some of them might be weak, irrelevant or hard to implement. Therefore, criteria are needed to evaluate each resulting idea.

Following an organizational prespective, the best ideas will be selected based on evaluation metrics and then stored in the form of ontology to enhance the accessibility and reusability of ideas. Results from previous studies (Boden 1998; Douglas et al. 2006; Jing et al. 2015), suggest that *quality, novelty, usefulness and surprisingness* are an appropriate set of criteria for evaluating ideas. The *quality* criterion can be measured by *Specificity* that measures the clarity of the idea, *feasibility, and relevance;* The usefulness criterion is measured by the sub-criteria of *Acceptability, Completeness, and Implicational Explicitness*. The idea could be evaluated as *surprising* if it is *unusual and unexpected*. Figure 5 shows the criteria used in the idea evaluation. *Novelty* has an importance weight and is given a value depending on the nature of the idea. For Example: If the idea is a new product, the novelty is determined by the existence of similar products in the market. Each criterion is given a weight based on its importance, This weight changes with the nature of the problem. Extending (Boden 1998; Douglas et al. 2006; Jing et al. 2015), the evaluation formula $C_i$ for an idea *i* can be calculated as:

$$C_i = \frac{N_i \times W_N + U_i \times W_U + Q_i \times W_Q + S_i \times W_s}{W_s + W_N + W_{U+W_Q}}$$

Where $N_i$, $W_N$ is the measure and weight of *Novelty*; $U_i$, $W_U$ is the value and weight of *Usefulness*; $Q_i$, $W_Q$ is the value and weight of *Quality*; and $S_i$, $W_s$ is the value and weight of *Surprising*.

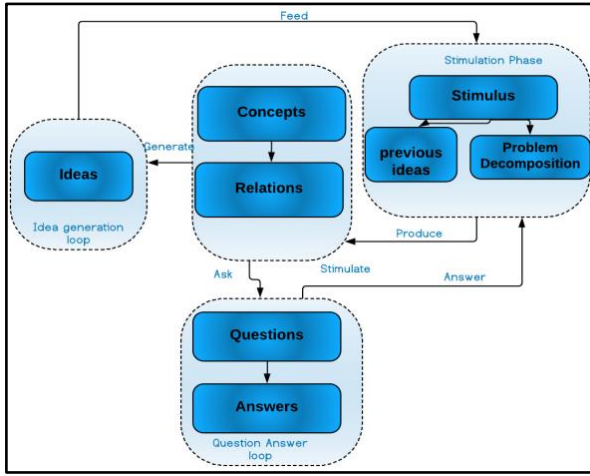
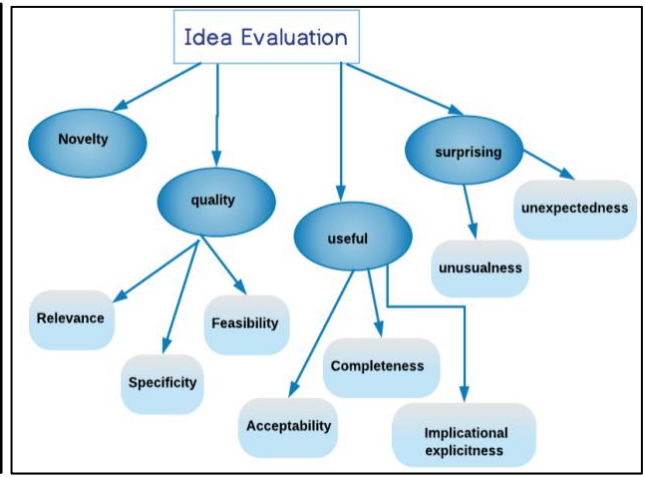

**Figure 4: Ideation Process Workflow**   **Figure 5: Ideas Evaluation Criteria**

### *Idea Ontology Generator*

After ideas are selected they are saved in the form of ontology to make easier for search and retrieval. The hierarchy of the Idea Ontology is shown in Figure 6. The ontology hierarchy is based on the contributing entities in the ideation session. Any generated idea is the output of the related *concepts and relations*. We also need to track which *departments* were engaged in developing the idea and store the *idea type* of the idea such as product, service or rule. In addition, information about the session's time, place, *context* or topics of the idea need to be saved in the *Session entity*. *Resources* that are needed to implement the idea, and the *questions* used as stimulus for generating the idea are also stored in the ontology.

## Implementation and Illustrative Example:

The framework was instantiated using the IBM Watson discovery, knowledge studio and the Github Watson discovery API to call and visualize Watson services. We choose the context of electronic cooking pots as an example. We gathered related data from different data files in different formats such as internet forums, Amazon reviews and different manufactures' products manuals. When new data enter to the KMS, the *Ideation Central* triggers the Watson *discovery service* to digest the data. The scenario for an ideation session is given below.

The company ABC is specialized in electronic cooking pots has started to define the ideation problem as "the company wants to add new features to the product XYZ". The group started the ideation session with preliminary ideas in mind and ask the framework interface to start an ideation session. Then, the *Workflow Manager* starts the ideation process workflow.

### *The first round*

**Stimulation phase:** The group decomposed the problem into the concepts of new *features* and the *drawbacks* of the current product. To gain more knowledge, the group started asking questions to the framework interface. The questions as sent to the QA system in the *IBM service layers*.

**Question Answering phase**:
- What are the latest technologies used in cooking pots by the rival companies?
- What are the latest technologies in market?

The *Ideation Central* sent the first question directly to be answered. However, the second question was too broad and retrieved no confident answer. So, the ideation central suggests questions in the same context from *Question Repository* such as "*What are the latest technologies in cooking pots?*", "*What are the latest technologies in electronic devices?*". The group choses the first question. Then, the framework responded with answers as shown in figure 8. These answers were visualized using the *visualization* module and the group tried to visualize the concepts as word cloud as shown in figure 7. Based on the generated answers, the group decided to start another ideation round.

### *The second round*

**Stimulation phase:** The group wanted more information about the current product flaws, so they tried to gain more insights by asking the framework more questions.

**Question answer phase**:
- What do people dislike about the pot?
- What is the ratio between negative and positive reviews?

The answers to these questions offers the group insights that the major negatives of the products are nonfunctional flaws as shown in Figure 9. So, they decided to stop the current ideation round and move to the idea generation phase.

### *Idea generation and evaluation*

Based on results, the group came up with two ideas. "*Adding a Bluetooth chip to the pot*" and adding "*heat meter inside the pot*". To select one of the two ideas, the *Ideation Central* asked the *Ideas Evaluation Module* to evaluate the ideas. The framework asked the group to determine the weights of each evaluation criteria for each idea. The group gave the *feasibility, novelty, and usefulness* criteria the highest weights for both ideas. However, the group needed more information about the potential cost. So, the *Ideation Central* asked discovery to retrieve information about the cost of the meter and the chip. After evaluation equation is applied, they decided that adding a *Bluetooth chip* to the pot is the most useful and novel idea.

Finally, the ideation session was stored in ontological form using *Idea Ontology Generator* based on figure 6. In addition to the idea ontology, the generated ideas, context, and evaluation results were saved in the *Ideas Repository*. Questions were stored according to their context in the *Questions Repository*.

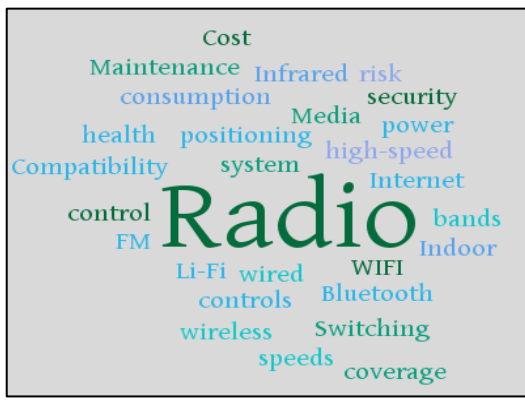
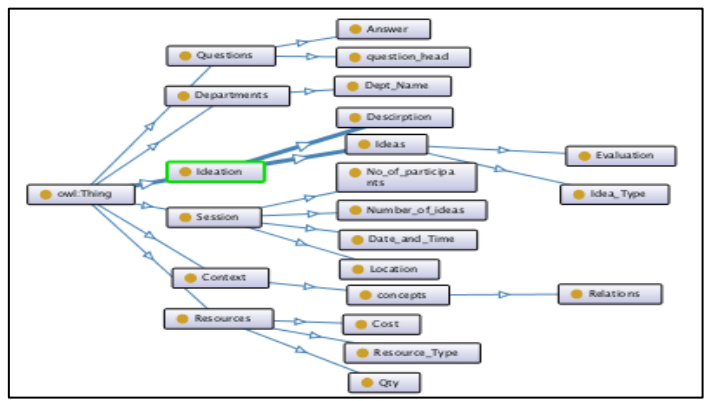

**Figure 7: Concept expansion visualization**     **Figure 6: Ideation Ontology**

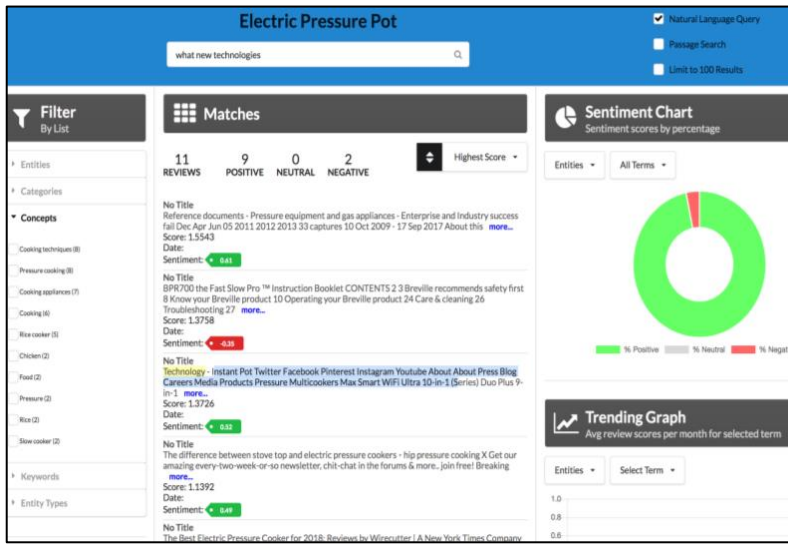
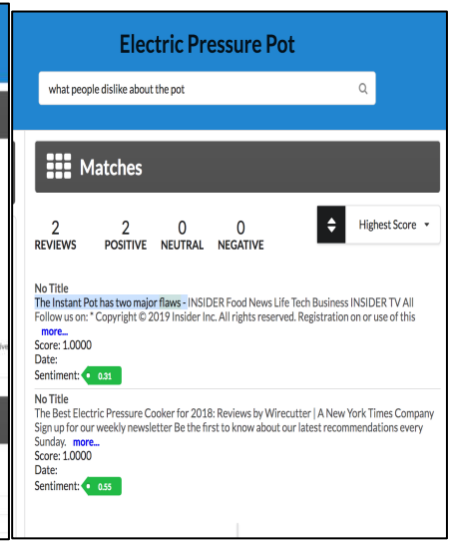

**Figure 8: framework results (round 1)**          **Figure 9: framework results (round 2)**

**Conclusion**

Idea generation is a core activity across a wide range of organizations and industries. However, the ideation process has many obstructions caused often by lack of knowledge, the limited cognitive abilities of individuals, and the ambiguous representation of the captured ideas. This paper aims to develop a framework that assist organizations in generating and storing ideas. The framework objective is to maximize the knowledge used in the ideation process through answering group members' questions. In addition, the framework creates concepts and relations that could be used as stimulus in the ideation process with the ability to visualize them at runtime. The framework uses IBM Watson cognitive services to answer questions, expand concepts and suggest ideas. At the end of the ideation session, the best ideas are evaluated and stored. Possibilities for future work addressing further issues such as security and accessibility, and also, the effects of psychological and social factors such as cognitive fixation, and evaluation apprehension.